\documentclass[fleqn,usenatbib,useAMS]{mnras}

\usepackage{newtxtext,newtxmath}
\usepackage[T1]{fontenc}

\DeclareRobustCommand{\VAN}[3]{#2}
\let\VANthebibliography\thebibliography
\def\thebibliography{\DeclareRobustCommand{\VAN}[3]{##3}\VANthebibliography}

\usepackage{graphicx}	% Including figure files
\usepackage{amsmath}	% Advanced maths commands
\usepackage{orcidlink}

\newcommand{\msun}{{\rm M}_{\sun}}

\title[Pollution in magnetic white dwarfs]{Sinking and spreading of metal pollution in magnetic white dwarfs}

\author[E. Shiftan and S. Ginzburg]{
Elad Shiftan\thanks{\raggedright E-mail: \href{mailto:elad.shiftan@mail.huji.ac.il}{elad.shiftan@mail.huji.ac.il} (ES);\newline
\href{mailto:sivan.ginzburg@mail.huji.ac.il}
{sivan.ginzburg@mail.huji.ac.il} (SG)}\label{emails} and
Sivan Ginzburg$^{\orcidlink{0000-0002-3751-4553}}$\hyperref[emails]{\footnotemark[1]}
\\
Racah Institute of Physics, The Hebrew University, Jerusalem 9190401, Israel
}

\date{Accepted XXX. Received YYY; in original form ZZZ}

\pubyear{\the\year{}}

\begin{document}
\label{firstpage}
\pagerange{\pageref{firstpage}--\pageref{lastpage}}
\maketitle

\begin{abstract}
Observations of magnetic polluted white dwarfs indicate that most of them have higher concentrations of metals near the poles compared to lower latitudes. Maintaining such abundance gradients requires gravitational sinking times $t_\downarrow$ below the convection zone that are shorter than the horizontal spreading time $t_\leftrightarrow$ across the surface by convective eddies. We show analytically that $t_\downarrow/t_\leftrightarrow\propto T^{8/3}$, where $T$ is the temperature at the base of the convection zone, which rises by more than an order of magnitude as convection penetrates deeper into the atmosphere. Correspondingly, $t_\downarrow/t_\leftrightarrow$ jumps by several orders of magnitude, clearly delineating between hot white dwarfs with abundance variations and cold ones with homogenous surfaces. We incorporate magnetic fields self-consistently into the stellar structure and compute for the first time the sinking to spreading time-scale ratio as a function of $B$. Magnetic white dwarfs have shallower convection zones at a given $T_{\rm eff}$, and the inward penetration of convection shifts to lower $T_{\rm eff}$.
Quantitatively, $B\sim 10^5\textrm{ G}$ extends the range for enhanced abundance patches from $T_{\rm eff}\gtrsim 30\,000 \textrm{ K}$ ($13\,000\textrm{ K}$) to $T_{\rm eff}\gtrsim 15\,000 \textrm{ K}$ ($6000\textrm{ K}$) for helium (hydrogen) dominated atmospheres, which is insufficient to explain the observed patchy white dwarfs, which are even colder. A possible solution is variability in the accretion rate on a time-scale of a decade, as recently detected in another white dwarf. 
\end{abstract}

\begin{keywords}
accretion, accretion discs -- convection -- stars: abundances -- stars: magnetic fields -- white dwarfs
\end{keywords}

%%%%%%%%%%%%%%%%%%%%%%%%%%%%%%%%%%%%%%%%%%%%%%%%%%

%%%%%%%%%%%%%%%%% BODY OF PAPER %%%%%%%%%%%%%%%%%%

\section{Introduction}

White dwarf photospheres are expected to consist of almost pure hydrogen or helium \citep[in case hydrogen was lost in a previous evolutionary phase; e.g.][]{Iben1983}, with heavier elements rapidly sinking inward under the influence of the strong gravity \citep{Schatzman1945, Schatzman1948, FontaineMichaud1979, Vauclair1979}. In practice, however, a large fraction of white dwarfs exhibit metal lines in their spectrum \citep{Zuckerman2003,Koester2014}, probably indicating atmospheric pollution by ongoing or recent accretion of disrupted minor planets or asteroids \citep{Jura2003,Jura2007,Zuckerman2010,Farihi2016}. These polluted white dwarfs provide a unique opportunity to probe the composition of extrasolar planetary systems \citep{JuraYoung2014,Harrison2021,Bonsor2023}.

Magnetism is another common feature in white dwarfs, and a spectropolarimetric survey found magnetic fields $B\sim 10^4-10^5$ G in four out of the thirteen polluted white dwarfs within 20 pc of the Sun \citep{BagnuloLandstreet2019}. Subsequent monitoring has revealed that in at least three of these stars, the magnetic field and the strength of the metal lines both change periodically as the star rotates, in synchronization with each other \citep{Bagnulo2024AA,Bagnulo2024ApJ,Bagnulo2026}. 
These variations have been interpreted as signs of magnetically guided accretion that produces patches of enhanced metal pollution at the poles.

The detection of metal abundance patches on the surface of these white dwarfs challenges our theoretical understanding. While magnetic fields as weak as $B\sim 10^3$ G may be sufficient to channel accretion to the magnetic poles \citep{GhoshLamb1978,Metzger2012,Farihi2018,Cunningham2021}, surface convection is expected to spread the accreted metals across the white dwarf's surface much faster than they sink below the convection zone. Specifically, \citet{Cunningham2021} compared the horizontal and vertical diffusion times, finding that horizontal spreading is faster than vertical sinking in white dwarfs cooler than $T_{\rm eff}\lesssim 10\,000$ K ($T_{\rm eff}\lesssim 30\,000$ K) with hydrogen (helium) dominated atmospheres. All the patchy polluted magnetic white dwarfs in the \citet{BagnuloLandstreet2019} sample have helium dominated atmospheres with $T_{\rm eff}\approx 6000-7000$ K, apparently defying this theoretical prediction.

A natural solution to this conundrum may be the magnetic field itself. \cite{Tremblay2015} demonstrated that sufficiently strong magnetic fields can inhibit convective energy transfer, potentially altering both the spreading and sinking time-scales \citep{Cunningham2021}. Recently, as part of a comprehensive analysis of magnetic white dwarf pollution, \citet{Pham2026} tried to quantify this effect by recalculating the ratio of the sinking and spreading time-scales in the presence of a magnetic field. However, as they emphasize in the paper, their analysis suffers from two shortcomings. First, \citet{Pham2026} adopted the horizontal diffusion coefficients of \citet{Cunningham2021}, which were derived from a non-magnetic white dwarf simulation, such that only the effect of the magnetic field on the vertical sinking time is considered. Second, they recalculated the sinking time only in the extreme limit where convection is artificially shut down entirely throughout the star, mimicking complete inhibition by a strong magnetic field. This limitation prevents them from associating their results with a specific magnetic field strength $B$. 
In addition to the sinking time, \citet{Pham2026} also considered the effect of magnetism on the convective turnover time, finding that magnetic fields $B\gtrsim 10^8$ G can change the convection zone's morphology.
Such strong fields may also slow down gravitational sinking by forcing the metal ions to gyrate around field lines in cyclotron motion \citep{MichaudFontaine1982,Pham2026}. 
These fields, however, are orders of magnitude stronger than observed for the patchy white dwarfs.    

Here, we attempt to fill these theoretical gaps and explain the recent observations by estimating both the sinking and spreading time-scales of metals in white dwarf atmospheres as a function of $B$. Specifically, we follow \citet{Ginzburg2024,Ginzburg2025} and incorporate magnetic fields self-consistently into the stellar structure using the \citet{GoughTayler1966} stiffened criterion for convective instability in the presence of magnetic fields.\footnote{Similarly to \citet{Tremblay2015}, \citet{GoughTayler1966} find that magnetic fields significantly affect convection only above a critical value $B^2\sim 8\upi P$. However, the pressure $P$ itself \citep[e.g. at the bottom of the convection zone; see][]{Ginzburg2024} changes, necessitating a self-consistent recalculation of the stellar structure with $B$ included, as done here.}  
This consistent treatment enables us to compute the structure of the white dwarf's convection zone as a function of $B$ and $T_{\rm eff}$, and derive from it the sinking and spreading time-scales. 

The remainder of this paper is organized as follows. In Sections \ref{sec:horizontal} and \ref{sec:vertical}, we explain our calculation of the horizontal and vertical time-scales, respectively, and provide some analytical intuition. In Section \ref{sec:magnetic}, we calculate how both of these time-scales change as a function of the magnetic field strength. 
We summarize and discuss our conclusions in Section \ref{sec:conclusions}.

\section{Horizontal spreading}\label{sec:horizontal}

When a white dwarf cools down sufficiently \citep[$T_{\rm eff}\lesssim 16\,000$ K for a $0.6\,\msun$ white dwarf with a hydrogen atmosphere; e.g.][]{BauerBildsten2019}, its outer layers recombine and a surface convection zone develops, gradually penetrating deeper as the white dwarf continues to cool \citep{Bohm1968,vanHorn1970,Fontaine2001}. Convective eddies spread the pollution horizontally in a random walk with an effective macroscopic diffusion coefficient
\begin{equation}\label{eq:mlt}
    D_\leftrightarrow\sim Hv,
\end{equation}
where $v$ is the typical eddy velocity and $H\sim kT/(mg)$ is its typical size, which in mixing-length theory is assumed to be similar to the scale height. We denote Boltzmann's constant by $k$, the surface gravity by $g$, and the dominant particle mass by $m$. Inside the convection zone, the temperature $T$ scales adiabatically with the density $T\propto\rho^{2/3}$, assuming a monoatomic ideal gas. 
The convective eddies also carry the flux $F$ outward, such that \citep[e.g.][]{Hansen2004}
\begin{equation}\label{eq:conv_flux}
    F=\sigma T_{\rm eff}^4\sim\rho v^3,
\end{equation}
where $\sigma$ is the Stefan--Boltzmann constant. The convection zone is negligible in both mass and radial thickness compared to the white dwarf, such that both $g$ and $F$ are uniform inside it. As a result, $v\propto \rho^{-1/3}$ in the convection zone, with the highest velocities reached close to the photosphere, where the density is lowest. Combining equations \eqref{eq:mlt} and \eqref{eq:conv_flux}, however, indicates that the diffusion coefficient reaches its maximum at the base of the convection zone, where the density is highest $D_\leftrightarrow\propto Tv\propto \rho^{2/3}\rho^{-1/3}\propto \rho^{1/3}$. We assume that the convection zone is fully mixed vertically, such that this maximal diffusion coefficient sets the horizontal spreading of metals at the surface. For the remainder of this section, $\rho$, $T$, and $D_\leftrightarrow$ are therefore evaluated at the base of the convection zone. 

For a simple analytical estimate of the density at the bottom of the convection zone, we follow \citet{Yaakovyan2025} and calculate the optical depth there $\tau$ assuming Kramers' opacity law $\kappa\propto\rho T^{-7/2}$. Using hydrostatic equilibrium
\begin{equation}\label{eq:tau_hs}
    \tau\sim\frac{\kappa P}{g}\propto \rho^2T^{-5/2},
\end{equation}
where $P\propto\rho T$ is the ideal gas pressure. At this transition between convection and radiation, the radiative diffusion equation is applicable, such that
\begin{equation}\label{eq:tau_rad}
    \tau\sim \frac{\sigma T^4}{F}=\left(\frac{T}{T_{\rm eff}}\right)^4.
\end{equation}
By comparing equations \eqref{eq:tau_hs} and \eqref{eq:tau_rad}, the density at the base of the convection zone scales as
\begin{equation}\label{eq:rho}
    \rho\propto T_{\rm eff}^{-2}T^{13/4},
\end{equation}
and using equations \eqref{eq:mlt} and \eqref{eq:conv_flux} the horizontal diffusion coefficient there scales as
\begin{equation}
    D_\leftrightarrow\propto T\left(\frac{T_{\rm eff}^4}{\rho}\right)^{1/3}\propto T_{\rm eff}^2T^{-1/12}.
\end{equation}
The horizontal diffusion coefficient defines a characteristic horizontal spreading time across the white dwarf's surface
\begin{equation}\label{eq:t_horizontal}
    t_\leftrightarrow\equiv\frac{R^2}{D_\leftrightarrow}\propto T_{\rm eff}^{-2}T^{1/12},
\end{equation}
where $R$ is the white dwarf's radius \citep[][up to order unity coefficients which we omit]{Cunningham2021,Pham2026}.

\citet{Yaakovyan2025} demonstrated that white dwarf convection zones evolve through three distinct phases as a function of $T_{\rm eff}$ \citep[see also][]{ChenHansen2011}: an initial phase in which $T$ increases from the recombination temperature by more than an order of magnitude \citep[accompanied, according to equation \ref{eq:rho}, by a sharp increase in $\rho$, and therefore in the mass of the convection zone; see][]{BauerBildsten2019}, an intermediate phase $6000\textrm{ K}\lesssim T_{\rm eff}\lesssim 10\,000\textrm{ K}$ (for a $0.6\,\msun$ white dwarf with a hydrogen atmosphere) during which $T\approx \textrm{const.}$, and a terminal phase when the convection zone is directly coupled to the thermally conducting degenerate core, such that equation \eqref{eq:tau_rad} is no longer applicable \citep{Fontaine2001}. Here we focus only on the first two phases (i.e. before convective coupling), during which we expect approximately $t_\leftrightarrow\propto T_{\rm eff}^{-2}$, up to a small jump in the initial phase due to the weak $\propto T^{1/12}$ dependence in equation \eqref{eq:t_horizontal}. The exact dependence on $T$ is actually more complicated, because the coefficient in Kramers' opacity law depends on the ionization level \citep{Hansen2004}. None the less, $t_\leftrightarrow\propto T_{\rm eff}^{-2}$ is a reasonable approximation for the intermediate stage in particular, during which $T$ is approximately constant.

\begin{figure}
\includegraphics[width=\columnwidth]{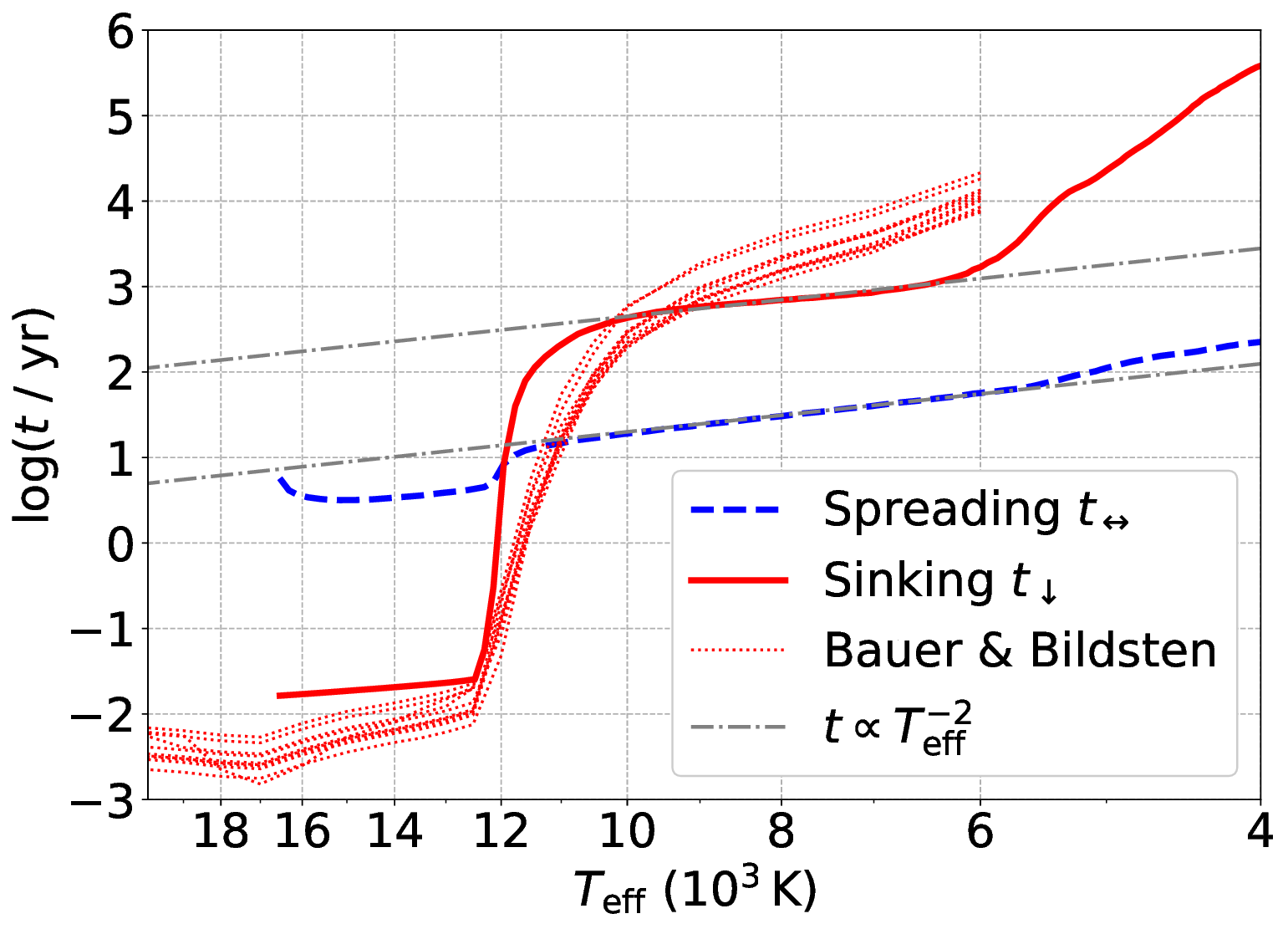}
\caption{Horizontal spreading ($t_\leftrightarrow$) and vertical sinking ($t_\downarrow$) time-scales for a non-magnetic $0.6\,\msun$ white dwarf with a hydrogen dominated atmosphere. $t_\leftrightarrow$ is computed using equation \eqref{eq:t_horizontal} with the maximal $D_\leftrightarrow$ in the convection zone. $t_\downarrow$ is estimated using equation \eqref{eq:t_vertical_H} with $H$, $\rho$, and $T$ evaluated at the base of the convection zone, and calibrated to fit \citet[][dotted red lines for the different elements in their table 2]{BauerBildsten2019}. From equations \eqref{eq:t_horizontal} and \eqref{eq:t_vertical} we expect that both time-scales are approximately $\propto T_{\rm eff}^{-2}$ for
$6000\textrm{ K}\lesssim T_{\rm eff}\lesssim 10\,000\textrm{ K}$.}
\label{fig:time_scales}
\end{figure}

In practice, in Fig. \ref{fig:time_scales} we compute the horizontal spreading time $t_\leftrightarrow$ more accurately using the test suite \texttt{wd\_cool\_0.6M} of the \textsc{mesa} stellar evolution code, version r23.05.1 \citep{Paxton2011,Paxton2013,Paxton2015,Paxton2018,Paxton2019,Jermyn2023}, by taking the maximal $D_\leftrightarrow$ in the convection zone (\textsc{mesa} computes $D_\leftrightarrow$ in each cell using mixing-length theory). We note that \citet{Cunningham2021} tested mixing-length theory by comparing equation \eqref{eq:mlt} to the actual diffusion of tracer densities in three-dimensional radiation hydrodynamics simulations of white dwarf atmospheres. \citet{Cunningham2021} focused on the diffusion coefficient close to the photosphere, where the convective velocity peaks $v\propto\rho^{-1/3}$. Here, on the other hand, we focus on the maximal diffusion coefficient, which peaks close to the base of the convection zone $D_\leftrightarrow\propto\rho^{1/3}$. Despite this and other differences, our computed diffusion coefficients $D_\leftrightarrow(T_{\rm eff})$ agree with \citet{Cunningham2021} up to a factor of a few for $T_{\rm eff}\lesssim 15\,000\textrm{ K}$, which is sufficient for analysing the competition between horizontal spreading and vertical sinking.  

\section{Vertical sinking}\label{sec:vertical}

A trace pollutant with mass and charge numbers $A_2$ and $Z_2$ respectively that is embedded in a dominant ambient species with $A_1$ and $Z_1$ will experience an effective gravity 
\begin{equation}\label{eq:geff}
    \tilde{g}=g\left(1-\frac{Z_2/A_2}{Z_1/A_1}\right),
\end{equation}
when taking into account the electrical field caused by the gravitational separation of electrons and ions \citep[see][an references therein]{BauerBildsten2019}. This effective gravity accelerates the polluting ions to a drift velocity $w$, at which it is balanced by collisions
\begin{equation}\label{eq:w_drift}
\begin{split}
    w\sim\frac{l}{v_{\rm th}}&\left[\tilde{g}\frac{A_2}{A_1}+\frac{1}{\rho}\frac{{\rm d}P}{{\rm d} r}\left(1-\frac{Z_2}{Z_1}\right)\right]\\
    \sim D_\downarrow &\left[\frac{m\tilde{g}}{kT}\frac{A_2}{A_1}+\frac{{\rm d}\ln P}{{\rm d}r}\left(1-\frac{Z_2}{Z_1}\right)\right],
    \end{split}
\end{equation}
where the second term is due to the ion pressure gradient ($r$ is the radial coordinate), which is also partially balanced by the electrical field \citep[see][for a complete derivation]{Pelletier1986}. The thermal velocity of the colliding particles (of the dominant species, assuming they have a much lighter mass) is given by $v_{\rm th}\sim (kT/m)^{1/2}$, we denote by $l$ their mean free path between collisions, and
\begin{equation}\label{eq:Diff_micro}
    D_\downarrow\sim v_{\rm th}l
\end{equation}
is the associated diffusion coefficient. The time it takes ions to sink below the convection zone, with a pressure scale height $H\equiv |{\rm d}\ln P/{\rm d}r|^{-1}\sim kT/(mg)$, is therefore
\begin{equation}\label{eq:t_down}
    t_\downarrow\sim\frac{H}{w}\sim \frac{H^2}{D_\downarrow},
\end{equation}
up to coefficients that depend on $A_{1,2}$ and $Z_{1,2}$. Equation \eqref{eq:t_down} indicates that gravitational settling operates on roughly the same time-scale as diffusion due to a concentration gradient; see equation (3) of \citet{BauerBildsten2019}, which includes all of these terms, in addition to a similar term of thermal diffusion.

In principle, the mean free path $l$ can be estimated from the impact parameter $b$ for strong collisions in the plasma that deflect the thermal particles significantly 
\begin{equation}
    l\sim\frac{1}{n\Sigma}\sim\frac{1}{nb^2}\sim \frac{1}{n}\left(\frac{Z_1Z_2e^2}{kT}\right)^{-2},
\end{equation}
where $n=\rho/m$ is the number density, $e$ is the electron charge, and $\Sigma$ is the effective cross section for collisions. Using equation \eqref{eq:Diff_micro}, the microscopic diffusion coefficient scales as
\begin{equation}\label{eq:D_down}
    D_\downarrow\sim\frac{(kT)^{5/2}}{nm^{1/2}Z_1^2Z_2^2e^4}\sim\frac{m^{1/2}(kT)^{5/2}}{\rho Z_1^2Z_2^2e^4}\propto\frac{T^{5/2}}{\rho}.
\end{equation}
However, it is well known that cumulative weak collisions contribute to the cross section as much as strong collisions, such that $\Sigma$ is multiplied by the Coulomb logarithm, which is truncated at the plasma's screening length \citep{Paquette1986,StantonMurillo2016}. 

We can analytically estimate the sinking time-scale using equations \eqref{eq:t_down} and \eqref{eq:D_down}, and the conditions ($H$, $\rho$, $T$) at the bottom of the convection zone
\begin{equation}\label{eq:t_vertical_H}
   t_\downarrow\propto \frac{H^2\rho}{T^{5/2}}.
\end{equation}
Substituting $H\propto T$ and the density $\rho$ from equation \eqref{eq:rho}
\begin{equation}\label{eq:t_vertical}
    t_\downarrow\propto \rho T^{-1/2}\propto T_{\rm eff}^{-2}T^{11/4} 
\end{equation}
demonstrates that, unlike $t_\leftrightarrow\propto T^{1/12}$, the sinking time lengthens dramatically as the base of the convection zone heats up (and deepens drastically) shortly after the onset of convection.  
Specifically, equations \eqref{eq:t_horizontal} and \eqref{eq:t_vertical} indicate that the ratio of the two time-scales
\begin{equation}\label{eq:t_ratio}
    \frac{t_\downarrow}{t_\leftrightarrow}\propto T^{8/3}
\end{equation}
increases by orders of magnitude during this stage, before reaching a plateau once $T$ saturates \citep[at $T_{\rm eff}\lesssim 10\,000\textrm{ K}$ for a hydrogen-atmosphere $0.6\,\msun$ white dwarf, see][]{Cunningham2021,Pham2026}.

An accurate calculation of the sinking time $t_\downarrow$ requires diffusion coefficients that include the Coulomb logarithm \citep{Paquette1986,Paquette1986b}, as well as the ionization level $Z_2$ of each polluting species \citep[][note that $Z_2$ also appears directly in their equation 3 and in our equations \ref{eq:geff} and \ref{eq:w_drift}]{Dupuis1992,BauerBildsten2019}. In practice, the sinking times of different elements are similar to within a factor of a few \citep{BauerBildsten2019,Cunningham2021}, such that we may define a single $t_\downarrow$ for metal pollution. We make a further simplification and adopt the approximate scaling of equation \eqref{eq:t_vertical_H}, which lacks the Coulomb logarithm and ionization level. We account for these factors, as well as additional order of unity coefficients that we have omitted, by calibrating the coefficient in front of the scaling to the accurate \textsc{mesa} computations of \citet{BauerBildsten2019}, which are in agreement with \citet{Koester2009}.

As demonstrated in Fig. \ref{fig:time_scales}, our simplified approach captures the qualitative behaviour of $t_\downarrow(T_{\rm eff})$, and specifically its sharp rise by about four orders of magnitude as $T$ at the base of the convection zone increases for $10\,000\textrm{ K}\lesssim T_{\rm eff}\lesssim 13\,000\textrm{ K}$, in accordance with equation \eqref{eq:t_vertical}. This sharp rise implies that $t_\downarrow/t_\leftrightarrow\ll 1$ for $T_{\rm eff}\gtrsim 13\,000\textrm{ K}$, whereas $t_\downarrow/t_\leftrightarrow\gg 1$ for $T_{\rm eff}\lesssim 10\,000\textrm{ K}$, such that metal abundance patches are not expected for such cool white dwarfs, in agreement with \citet{Cunningham2021} and \citet{Pham2026}. Quantitatively, our approximate $t_\downarrow$ is within an order of magnitude of \citet{BauerBildsten2019} for most of the $T_{\rm eff}$ range. As we show in Section \ref{sec:magnetic}, our method is thus sufficiently accurate to study the effects of the magnetic field $B$ on the time-scale ratio $t_\downarrow/t_\leftrightarrow$, while remaining analytically tractable (e.g. equation \ref{eq:t_ratio}). In the future, the accuracy can be improved further -- at the expense of computational simplicity and flexibility -- by injecting metals at a constant rate in \textsc{mesa} for all combinations of $T_{\rm eff}$ and $B$, and measuring their diffusion velocity once a steady state is reached \citep{BauerBildsten2019,Pham2026}.  

\section{Magnetic fields}\label{sec:magnetic}

We incorporate magnetic fields into \textsc{mesa} in the same manner as \cite{Ginzburg2025}, by using the \texttt{other\_mlt\_results} hook and setting the temperature gradient in each cell $\nabla\equiv {\rm d}\,\ln T/{\rm d}\,\ln P$ to
\begin{equation}\label{eq:gough_mesa}
\nabla=\min\left(\nabla_{\rm rad},\nabla_{\rm ad}+\frac{B^2}{B^2+4\upi P}\right),    
\end{equation}
where $\nabla_{\rm rad}$ and $\nabla_{\rm ad}$ are the radiative and adiabatic temperature gradients, respectively, $P$ is the pressure, and $B$ is the magnetic field. Equation \eqref{eq:gough_mesa} follows the \citet{GoughTayler1966} criterion for convection, which reduces to the standard Schwarzschild criterion for $B=0$, and partially inhibits convection for $B>0$. We have omitted from the criterion the adiabatic index $\gamma\equiv 1/(1-\nabla_{\rm ad})\sim 1$, which \citet{GoughTayler1966} treat arbitrarily, and which does not impact our conclusions \citep[see also][]{MullanMacDonald2001}. This method enables us to compute self-consistent white dwarf profiles and convection zones for any given $T_{\rm eff}$ and $B$, improving upon \citet{Pham2026} who turned convection off for the entire star.

\begin{figure}
\includegraphics[width=\columnwidth]{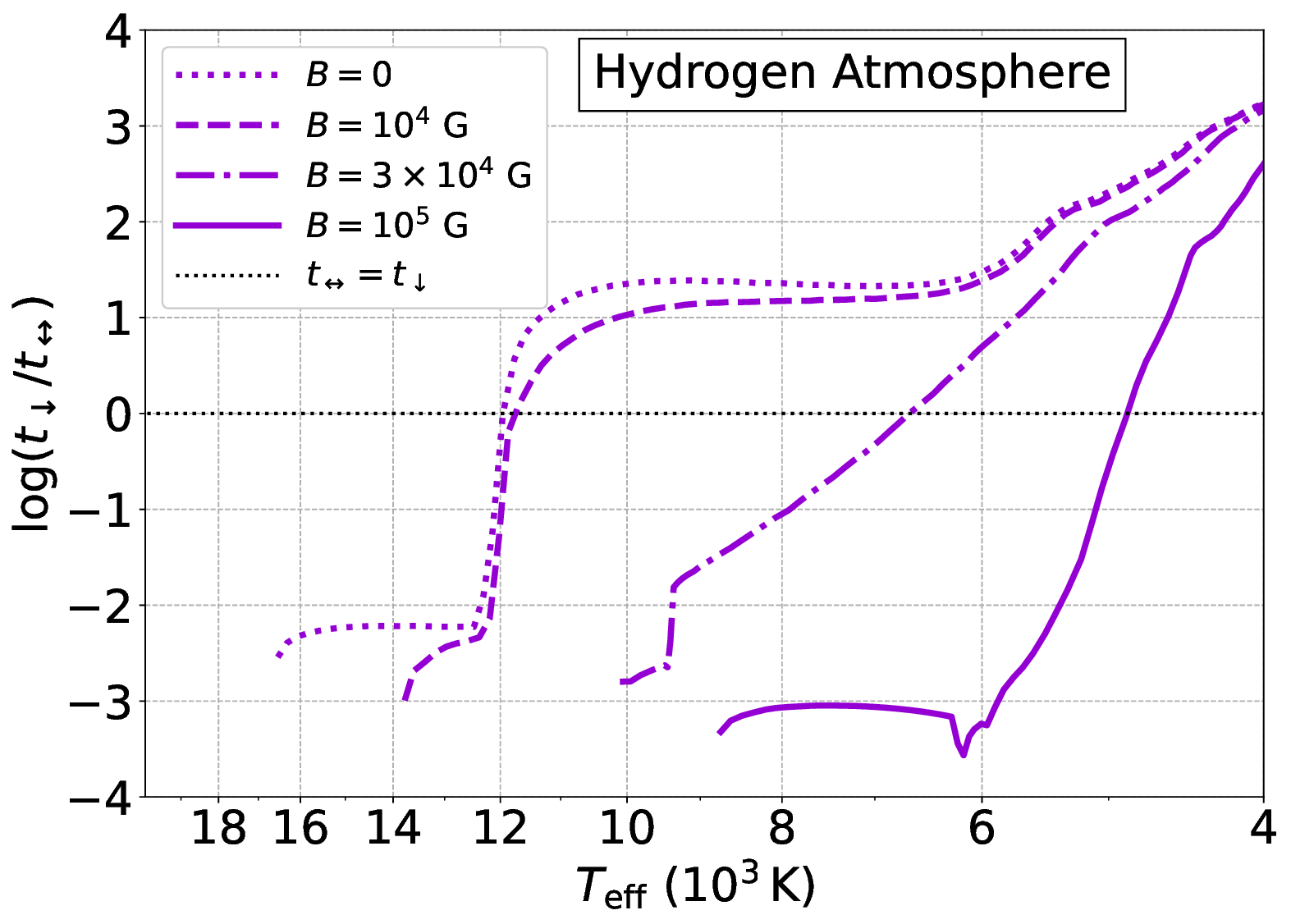}
\caption{The ratio of sinking to spreading time-scales $t_\downarrow/t_\leftrightarrow$ as a function of $T_{\rm eff}$ and the magnetic field $B$ for a $0.6\,\msun$ white dwarf with a hydrogen dominated atmosphere. Both time-scales are computed self-consistently for magnetic white dwarfs by replacing the Schwarzschild criterion for convection with \citet{GoughTayler1966}. For $B=0$, the time-scales in Fig. \ref{fig:time_scales} are reproduced.}
\label{fig:magnetic}
\end{figure}

We present the effect of $B$ on the sinking to spreading time-scale ratio $t_\downarrow/t_\leftrightarrow$ in Fig. \ref{fig:magnetic} (unlike \citealt{Pham2026}, we compute how both time-scales change as a function of $B$).
Similarly to \citet{Ginzburg2025}, we find that at a given $T_{\rm eff}$, stronger magnetic fields yield shallower convection zones. Moreover, magnetic fields delay the inward penetration of the convection zone, such that its base reaches high densities and temperatures only at much lower $T_{\rm eff}$. As explained analytically in Section \ref{sec:vertical}, this delayed penetration of the convection directly shifts the jump in $t_\downarrow/t_\leftrightarrow$ to lower $T_{\rm eff}$. Quantitatively, Fig. \ref{fig:magnetic} indicates that magnetic fields $B\sim 10^5\textrm{ G}$ extend the range where metal abundance patches are possible, i.e. $t_\downarrow/t_\leftrightarrow\ll 1$, to $T_{\rm eff}\gtrsim 6000\textrm{ K}$ for a hydrogen-atmosphere $0.6\,\msun$ white dwarf (compared to $T_{\rm eff}\gtrsim 13\,000\textrm{ K}$ for $B=0$). 

\begin{figure}
\includegraphics[width=\columnwidth]{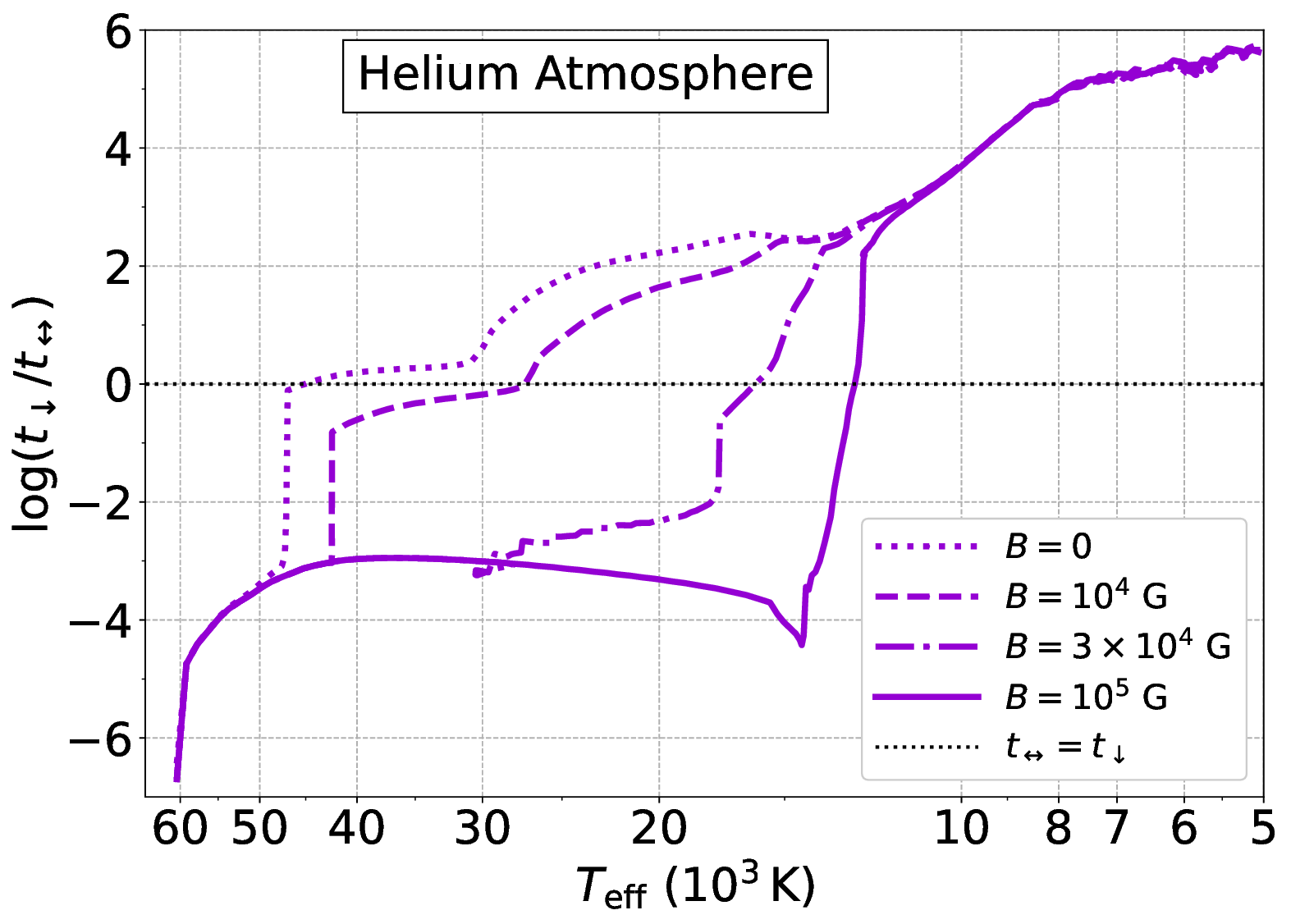}
\caption{Same as Fig. \ref{fig:magnetic}, but for a helium dominated atmosphere.}
\label{fig:helium}
\end{figure}

In Fig. \ref{fig:helium}, we repeat our calculations for helium dominated atmospheres, which develop convection at a much higher $T_{\rm eff}$ \citep[e.g.][]{Bedard2024}.\footnote{None the less, we may neglect radiative levitation \citep[see][]{Chayer1995}, because $B$ shifts the transition point $t_\downarrow\sim t_\leftrightarrow$ to lower $T_{\rm eff}$.} The effect of magnetic fields is similar to the hydrogen case, with $B\sim 10^5\textrm{ G}$ extending the possibility for patchy atmospheres to $T_{\rm eff}\gtrsim 15\,000\textrm{ K}$ (compared to at least $T_{\rm eff}\gtrsim 30\,000\textrm{ K}$ for $B=0$).
Interestingly, this is insufficient to explain the recently discovered patchy polluted white dwarfs, which have helium dominated atmospheres, $B\sim 10^4-10^5\textrm{ G}$ magnetic fields, but even lower $T_{\rm eff}\approx 6000-7000\textrm{ K}$ \citep{BagnuloLandstreet2019,Bagnulo2024AA,Bagnulo2024ApJ,Bagnulo2026}. 

None the less, our results might be applicable for explaining the persistence \citep[but not necessarily the formation; see][]{Ginzburg2025} of inhomogeneities in another emerging class of double-faced white dwarfs, which exhibit variations in their hydrogen and helium lines as they rotate \citep{Achilleos1992,Caiazzo2023,Moss2025}. Some of these stars have $15\,000\textrm{ K}\lesssim T_{\rm eff}\lesssim 30\,000\textrm{ K}$ and measured magnetic fields $B\gtrsim 10^6\textrm{ G}$, such that $t_\downarrow/t_\leftrightarrow\ll 1$ and hydrogen may float to the top of a helium convection zone without spreading horizontally across the surface (more generally, the thickness of the hydrogen layer may remain different). See \citet{BedardTremblay2025} for a comprehensive discussion of different possibilities for the prototype of this class.

We note that for helium atmospheres, our results for $B=0$ show some minor differences compared to previous studies \citep{Cunningham2021,Pham2026} due to the atmospheric boundary conditions and mixing-length parameter choice \citep[see][for the sensitivity to this parameter in particular]{Ginzburg2025,Pham2026}. In addition, using the same calibration \citep{BauerBildsten2019} for hydrogen and helium atmospheres introduces another order of unity deviation.
These minor differences do not affect our conclusions.

\section{Conclusions and discussion}\label{sec:conclusions}

Monitoring of the polluted magnetic white dwarfs in the local 20 pc sample has revealed that most (and perhaps all) of them have enhanced metal abundances around their magnetic poles compared to lower latitudes \citep{Bagnulo2024AA,Bagnulo2024ApJ,Bagnulo2026}. Maintaining such abundance gradients requires vertical gravitational metal sinking times $t_\downarrow$ that are shorter than the horizontal spreading time $t_\leftrightarrow$ by convective eddies across the white dwarf's surface.\footnote{\citet{Pham2026} distinguished between metal abundance patches ($t_\downarrow\ll t_\leftrightarrow$) and gradients ($t_\downarrow\sim t_\leftrightarrow$). We make no such distinction, and use the two terms interchangeably.} 

We computed these time-scales for non-magnetic white dwarfs using the \textsc{mesa} stellar evolution code, reproducing previous conclusions that $t_\downarrow \ll t_\leftrightarrow$ only as long as the convection zone is very shallow or absent, i.e. $T_{\rm eff}\gtrsim 13\,000\textrm{ K}$ ($T_{\rm eff}\gtrsim 30\,000\textrm{ K}$ at least) for a $0.6\,\msun$ white dwarf with a hydrogen (helium) atmosphere \citep{Cunningham2021,Pham2026}. Specifically, using analytical approximations for the microscopic and macroscopic (i.e. convective) diffusion coefficients, we found that $t_\downarrow/t_\leftrightarrow\propto T^{8/3}$, where $T$ is the temperature at the base of the confection zone, which rises sharply by more than an order of magnitude as the convection zone deepens \citep{Koester2009,ChenHansen2011}. This strong scaling implies a jump of several orders of magnitude in the time-scale ratio, clearly delineating between hot (in terms of $T_{\rm eff}$) white dwarfs with $t_\downarrow \ll t_\leftrightarrow$, where abundance gradients are expected, and cold ones with $t_\downarrow \gg t_\leftrightarrow$, which are supposed to have homogenous surfaces. 

Then, by incorporating the \citet{GoughTayler1966} criterion for convection into \textsc{mesa}, we computed for the first time $t_\downarrow/t_\leftrightarrow$ as a function of the magnetic field $B$. Magnetic fields stiffen the criterion for convective instability, such that at a given $T_{\rm eff}$ the convection zone is shallower, and its inward penetration shifts to lower $T_{\rm eff}$ \citep{Ginzburg2025}. Quantitatively, we find that $B\sim 10^5\textrm{ G}$ extends the range where abundance gradients are expected to $T_{\rm eff}\gtrsim 6000\textrm{ K}$ ($T_{\rm eff}\gtrsim 15\,000\textrm{ K}$) for hydrogen (helium) atmospheres. This result is much tighter than the limiting case of no convection at all \citep{Pham2026}, such that the recently detected abundance gradients (in helium atmospheres with $T_{\rm eff}\approx 6000-7000\textrm{ K}$ and $B\sim 10^4-10^5\textrm{ G}$) remain unexplained. 

While our results may be sensitive to the details of convective overshoot \citep{Cunningham2019}, it would only deepen the mixed region at a given $T_{\rm eff}$, resulting in an even higher $t_\downarrow/t_\leftrightarrow$. Thermohaline mixing is unimportant for helium atmospheres at such $T_{\rm eff}$ \citep{BauerBildsten2019}. Any other additional mixing mechanism would again only increase $t_\downarrow/t_\leftrightarrow$, either by deepening the mixed region (vertical mixing), or by directly shortening $t_\leftrightarrow$ (horizontal mixing). \citet{Stevenson1979} suggested that the convective eddy size and velocity decrease gradually as a function of $B$ inside the convection zone \citep[see also][]{BessilaMathis2024}, potentially lengthening $t_\leftrightarrow$ according to equations \eqref{eq:mlt} and \eqref{eq:t_horizontal}. However, since we assume that horizontal spreading is dominated by the
maximal $D_\leftrightarrow \sim Hv$ in the convection zone, we do not expect this effect to significantly reduce $t_\downarrow/t_\leftrightarrow$ beyond the sharp criterion of \citet{GoughTayler1966}.

One potential resolution of the cold patchy white dwarfs conundrum \citep[see also][]{Pham2026} is variability in the accretion rate on a time-scale that is shorter than $t_\leftrightarrow\sim 10^2\textrm{ yr}$, but longer than the white dwarf's rotation period, which is up to 1 yr \citep{Bagnulo2026}. This possibility could be tested observationally by further monitoring of the patchy polluted white dwarfs in the coming decade. In this context, see \citet{Farihi2026}, who recently reported for the first time changes in the accretion rate of a polluted white dwarf on a 25 yr time-scale.

Although our method incorporates magnetic fields self-consistently into the stellar structure and evolution, it is limited by its one-dimensional nature. Convection is inherently a three-dimensional phenomenon, especially in the presence of magnetic fields and rotation. Accurately analysing the interplay between convection and magnetic fields may therefore require a suite of three-dimensional magnetohydrodynamic simulations of white dwarf atmospheres \citep[e.g. extending the work of][]{Tremblay2015}. Our current work highlights the relevant range of $T_{\rm eff}$ and $B$ that should be explored by such simulations and by observations.   

\section*{Acknowledgements}

This research was partially supported by the United States-Israel Binational Science Foundation (BSF; grant no. 2022175), the German-Israeli Foundation for Scientific Research and Development (GIF; grant no. I-1567-303.5-2024), and the Israel Science Foundation (ISF; grant nos 1600/24 and 1965/24).

%%%%%%%%%%%%%%%%%%%%%%%%%%%%%%%%%%%%%%%%%%%%%%%%%%
\section*{Data Availability}
 
The data underlying this article will be shared on reasonable request to the corresponding authors.



\bibliographystyle{mnras}
\bibliography{scars} % if your bibtex file is called example.bib

%\begin{thebibliography}{99}
%\end{thebibliography}

%%%%%%%%%%%%%%%%%%%%%%%%%%%%%%%%%%%%%%%%%%%%%%%%%%

%%%%%%%%%%%%%%%%% APPENDICES %%%%%%%%%%%%%%%%%%%%%

%\appendix


%%%%%%%%%%%%%%%%%%%%%%%%%%%%%%%%%%%%%%%%%%%%%%%%%%

% Don't change these lines
\bsp	% typesetting comment
\label{lastpage}
\end{document}